\documentclass[review]{elsarticle}

\usepackage{color}
\usepackage{graphicx}
\usepackage{caption}
\usepackage{subcaption}
\usepackage{amsmath,amsfonts}

\newtheorem{lemma}{Lemma}[section]

\newtheorem{assumption}{Assumption}[section]
\newtheorem{corollary}{Corollary}[section]

\newtheorem{definition}{Definition}[section]

\journal{Journal of Nano Communication Networks}

\begin{document}

\begin{frontmatter}

\title{Channel Impulse Response-based Source Localization in a Diffusion-based Molecular Communication System}

\author{Henry Ernest Baidoo-Williams\textsuperscript{1,*}}
\author{Muhammad Mahboob Ur Rahman\textsuperscript{2,*}}
\author{Qammer Hussain Abbasi\textsuperscript{3}}
\address{\textsuperscript{1}Amazon Inc., USA}
\address{\textsuperscript{2}Electrical Engineering department, Information Technology University, Lahore, Pakistan}
\address{\textsuperscript{3}Department of Electronics and Nano engineering, University of Glasgow, Glasgow, UK}
\address{\textsuperscript{*}(Equal contribution by the authors)}




\begin{abstract}
This work localizes a molecular source in a diffusion based molecular communication (DbMC) system via a set of passive sensors and a fusion center. Molecular source localization finds its applications in future healthcare systems, including proactive diagnostics. In this paper, we propose two distinct methods which both utilize (the peak of) the channel impulse response measurements to uniquely localize the source, under assumption that the molecular source of interest lies within the open convex-hull of the sensor/anchor nodes.  The first method is a one-shot, triangulation-based approach which estimates the unknown location of the molecular source using least-squares method. The corresponding Cramer-Rao bound (CRB) is also derived. The second method is an iterative approach, which utilizes gradient descent law to minimize a non-convex cost function. Simulation results reveal that the triangulation-based method performs very close to the CRB, for any given signal-to-noise ratio. Additionally, the gradient descent-based method converges to the true optima/source location uniformly (in less than hundred iterations).
\end{abstract}


\end{frontmatter}


\section{Introduction}\label{introduction}

A nano-scale, molecular communication system consists of a nano-transmitter (emitter) and nano-receiver (ligand-receptor) which are apart by a few micro-meters; information transfer between them is realized via exchange of molecules. In a {\it diffusion based} molecular communication (DbMC) system, molecules undertake a brownian motion governed by the diffusion process. The very slow diffusion of molecules through the fluid medium implies that the DbMC channel is a low-rate, broadcast channel \cite{Ian:TIT:2013}. DbMC, though still in infancy, is an emerging paradigm as it enables communication between autonomous bionano-machines which are in turn amicable for interfacing with the biological systems, say, inside the human body \cite{Nakano:TMBMC:2017},\cite{Wang:Chinacomm:2017}. DbMC has the potential to revolutionize the healthcare system; additionally, it finds its applications in environmental monitoring and military scenarios \cite{Ian:CN:2008},\cite{Nakano:TMBMC:2017}. As of today, the researchers have done the noise analysis \cite{Llatser:JSAC:2013}, computed the channel capacity \cite{Ian:TIT:2013}, designed modulation schemes \cite{Kuran:ICC:2011} and optimal receivers \cite{Llatser:JSAC:2013}, and much more (see the survey article \cite{Wang:Chinacomm:2017} which provides a comprehensive overview of the recent development in the field).

Source localization, on the other hand, is the umbrella term for a handful of techniques which locate a signal source by utilizing the measurements collected by a set of sensor/anchor nodes (at known locations). Source localization has been extensively studied to localize a radio-frequency signal source \cite{Sayed:SPMag:2005}, optical source \cite{Armstrong:CommMag:2013}, acoustic source \cite{KUNDU:Ultrasonics4:2014}, and radioactive source \cite{HBW:SPL:2013}. Most of the localization algorithms comprise of the following two steps: i) the sensor nodes construct some measurement (received signal strength, time of arrival, time difference of arrival, pathloss, distance) from the signal received from the signal source of interest, ii) the fusion center fuses the measurements collected by the sensor nodes to minimize an appropriate cost function. To this end, various techniques have been reported in the literature, e.g., semidefinite programming \cite{Dasgupta:SPL:2008},\cite{Dasgupta:TSP:2011}, second order cone programming, gradient descent \cite{HBW:SPL:2013}, weighted least-squares etc. 

Very recently, the researchers have started to investigate the distance estimation methods\footnote{Note that a number of algorithms reported in the literature on localization of a wireless source build upon the distance estimates obtained by the sensor nodes (see \cite{Sayed:SPMag:2005}, \cite{Dasgupta:TSP:2011} and the references therein).} and the performance bounds in DbMC systems \cite{Moore:TSP:2012,Wang:CommL:2015,Huang:GC:2013,Noel:GC:2014}. In \cite{Moore:TSP:2012}, authors propose to estimate the round-trip time and signal attenuation from the received feedback signal to in turn estimate the distance. \cite{Wang:CommL:2015} investigates two distance estimation methods based on the peak and energy of the concentration of the received molecules. Huang et. al. in \cite{Huang:GC:2013} do synchronization-free distance estimation using one-way signaling (peak concentration and double-spike method). \cite{Noel:GC:2014} computes the Cramer-Rao bound (CRB) for distance estimation in a DbMC channel. 

Another set of works broadly relevant to the scope of our paper is \cite{Nakano:ICC:2014,Okaie:TNB:2014,Okaie:ICNCC:2015,Iwasaki:Globecom:2016,Nakano:SPAWC:2016,Nakano:TC:2017,Yang:ICSCN:2018} where detection and tracking of a bionano target is performed using self-organizing, mobile bionano-sensors that are capable of releasing attractant and repellent molecules. Specifically, \cite{Nakano:ICC:2014} develops a partial differential equations based mathematical model for the target tracking problem. \cite{Okaie:TNB:2014} performs target tracking with the aim of targeted drug delivery. \cite{Okaie:ICNCC:2015} carries out in-silico experiments by utilizing chemotactic bacteria and provide some information-theoretic insights on the performance of their proposed target tracking scheme. \cite{Iwasaki:Globecom:2016} extends the previous works to track multiple targets. \cite{Nakano:SPAWC:2016,Nakano:TC:2017} propose a leader-follower model for target tracking, describe the model mathematically and estimate the model parameters via maximum-likelihood approach. Finally, \cite{Yang:ICSCN:2018} extends \cite{Okaie:ICNCC:2015} by utilizing relay nodes for increased chemotactic efficiency. However, contrary to the works \cite{Nakano:ICC:2014,Okaie:TNB:2014,Okaie:ICNCC:2015,Iwasaki:Globecom:2016,Nakano:SPAWC:2016,Nakano:TC:2017,Yang:ICSCN:2018} which rely solely upon chemical interactions, our work does the localization by leveraging a set of passive, static, sensor nodes which record the measurements and a fusion center that is capable of fusing them to perform the computations. 

On a side note, \cite{Giaretta:TIFS:2016} studies the vulnerability of the attractant/repellent-based bionano target localization methods to the Sentry attack and Blackhole attack, and provides Baye's rule and threshold approach based countermeasures.

{\bf Contributions and Outlook.} We propose two novel methods to do source localization in a DbMC system, namely, triangulation/least-squares-based method, and gradient descent-based method. For the triangulation-based method, we also compute the corresponding CRB. Some futuristic applications in the healthcare domain that could potentially benefit from the proposed method include early disease (e.g., cancer) detection, targeted drug delivery \cite{Okaie:TNB:2014},\cite{Raz:TNB:2015}, and quick toxicity detection.  
 
{\bf Outline.} The rest of this paper is organized as follows. Section II introduces the system model and the DbMC channel model. Section III describes the measurement model used by the two proposed methods. Section IV presents the two proposed methods for source localization. Section V provides some simulation results. Section VI concludes the paper.

\section{System Model \& Channel Model}
\label{sec:sys-model}

\subsection{System Model}
\begin{figure}[ht]
\begin{center}
	\includegraphics[width=3in]{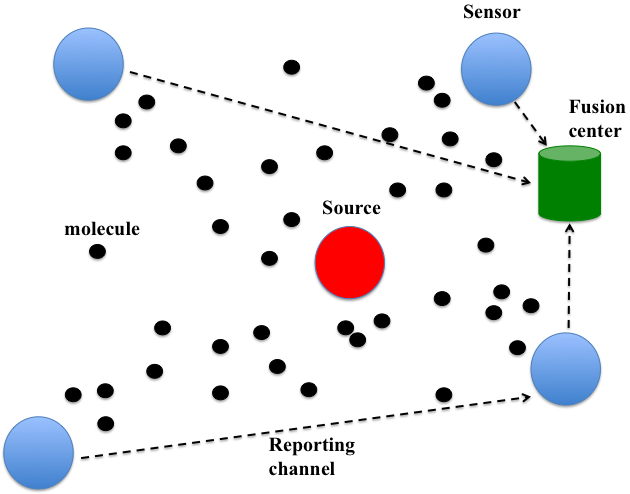} 
\caption{System model: The molecular source that is to-be localized lies within the convex hull of the nano sensor/anchor nodes (each of which receives the molecules emitted by the source). Moreover, the sensor nodes report their measurements to a fusion center which ultimately does the signal processing to localize the source.}
\label{fig:sysmodel}
\end{center}
\end{figure}

Consider a molecular source/transmitter, whose location ${y^*} \in \mathbb{R}^N$ (where $N\in \{2,3\}$) is to be estimated (see Fig. \ref{fig:sysmodel}). Source localization is traditionally done by deploying a set of sensor nodes in close vicinity of the source which report their measurements to a fusion center (FC). The FC is assumed to be a powerful node that is capable of performing sophisticated signal processing operations (therefore, FC is likely to be an on-body node). As for the sensor nodes, Triangulation based methods have shown that we need at least $n=N+1$ sensors located at ${x_i} \in \mathbb{R}^N, i \in \{1, \cdots, n\}$ which are non-collinear for $N=2$ case and non-coplanar for $N=3$ case. To keep the analysis tractable, this work assumes that: i) the sensors are passive receivers \cite{Noel:Globecom:2016}, ii) there is no interference caused by the molecules sent in previous slots, and iii) the reporting channel (i.e., the link between the sensor nodes and the FC) is error-free and delay-free.

Define $d_i$ as the euclidean distance between the source and $i$-th sensor/anchor node (with known location ${x_i}$):
\begin{equation} \label{eq:di}
	d_i = || {x_i} - {y^*} || 
\end{equation}
where $||.||$ is the 2-norm operator. 

Next, some assumptions and definitions.

\begin{assumption} \label{assumption:ass1}
The location of the source node $y^* \in \mathbb{R}^N$ is within the open convex hull of the measurement sensors. 
\end{assumption}
\begin{definition} \label{definition:d1}
Define ${\bf X} = [x_1, x_2, \cdots, x_n ]^T$.
\end{definition}

\begin{definition} \label{definition:d1}
Define ${\bf 1_n} = [1, \cdots, 1 ]^T$, and $n \times 1$-- dimensional vector with all entries being 1.
\end{definition}

\begin{corollary} \label{corollary: c1}
Under Assumption $\ref{assumption:ass1}$, $\exists \beta = [ \beta_1, \cdots, \beta_n]^T$, $\beta_i >0$ $\forall i \in \{1, \cdots, n\}$ such that $ {\bf 1_n}^T \beta=1$. Then,  
\begin{equation} \label{equations:e1}
 y^* =  {\bf X}^T \beta.
\end{equation}
\end{corollary}

Note that assumption \ref{assumption:ass1} can easily be satisfied using coarse initial estimates by perturbing the locations of the measurement sensors.

\subsection{The DbMC Channel Model}
Consider a DbMC system whereby the transmitter uses pulse-based modulation (i.e., on/off keying) and sends $Q$ molecules within one pulse. Since communication (transport of molecules from source to sensors) takes place through a diffusion paradigm, the DbMC channel can be described as broadcast channel, hence, the transmitter's ``message" can be received by all the sensors. Consequently, we can use Fick's second law of diffusion to characterize the mean change in concentration of molecules at a fixed distance $d_i$ w.r.t. time (because diffusion is a stochastic process):
\begin{equation} \label{eq:Fick_eq}
	\frac{\partial p(d_i,t | d_0)}{\partial t} = D \nabla^2 p(d_i,t | d_0)
\end{equation}
where $\nabla^2$ is the Laplacian operator, $p(d_i,t | d_0)$ is the molecule distribution function at time $t$, distance $d_i$ given the initial distance $d_0$, and $D$ is diffusion coefficient of the medium.

The solution to (\ref{eq:Fick_eq}), given in (\ref{eq:Fick_sol}), is the expected concentration of molecules as a function of time and distance (which is also the impulse response of the DbMC channel), where $c_i (d_i,t)$ denotes the concentration at distance $d_i$ and time $t$ from the the initial transmission time: 
\begin{equation} \label{eq:Fick_sol}
	c_i (d_i,t) = \frac{Q}{(4\pi Dt)^{\frac{3}{2}}} e^{-\frac{d_i^2}{4Dt}} 
\end{equation}

\begin{figure}[ht]
\begin{center}
	\includegraphics[width=3.5in]{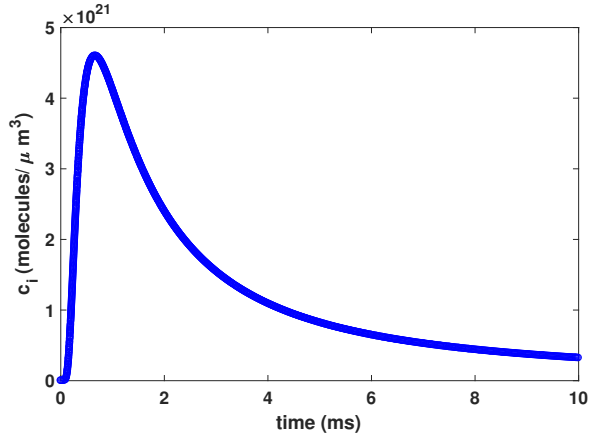} 
\caption{The received molecular pulse at $i$-th sensor (for $Q=5\times 10^5$, $D=1e-9$ m$^2$/sec): the blue curve represents the ideal impulse response at $d_i=2$ $\mu$m dictated by (\ref{eq:Fick_sol}), while the red dots represent actual/noisy measurements made by the $i$-th sensor.}
\label{fig:channel-model}
\end{center}
\end{figure}

A typical pulse following the model in (\ref{eq:Fick_sol}), as seen by the $i$-th sensor is shown in Fig. \ref{fig:channel-model}.

\section{Measurement Model for Source Localization}
\label{sec:methods}

Assume that $M$ measurements $z_i[k]$ ($k=1,...,M$) of channel impulse response (CIR) are taken by sensor $i$ during a single observation interval.
\begin{equation}
	\label{eq:rss1}
	z_i[k] = c_i (d_i,k) + \omega_i[k] = \frac{Q}{(4\pi Dk T_s)^{\frac{3}{2}}} e^{-\frac{d_i^2}{4Dk T_s}} + \omega_i[k]
\end{equation}
where $\omega_i[k]$ is the Poisson noise with independent and identically distributed (i.i.d.) elements, and $T_s$ is the sampling period of the system. In this work, sensor $i$ picks the largest measurement $z_i$ (corresponding to the instant where received molecular concentration was maximum)\footnote{The peak of CIR is analogous to the notion of received signal strength in wireless communication.} as:
\begin{equation}
	\label{eq:rss2}
	z_i = c_{max,i}(d_i) + \omega_i = \left(\frac{3}{2\pi D e^{\frac{1}{D}}}\right)^{\frac{3}{2}} \frac{Q}{d_i^3} + \omega_i
\end{equation}
and sends it to the fusion center. {\color{black} Here $\omega_i$ is distributed as Poisson random variable with parameter $\lambda=\frac{\alpha}{d_i^3}$ where $\alpha = \left(\frac{3}{2\pi D e^{\frac{1}{D}}}\right)^{\frac{3}{2}} Q$.}

\section{Source Localization}
\label{sec:current-wireless}
In the sequel, we make the following assumption:
\begin{assumption} \label{assumption:ass2}
The location of the sensor $x_i, i \in \{1, \cdots, n\}$ is not coincident with the location of the source, $y^*$. 
\end{assumption}
Assumption \ref{assumption:ass2} makes sense since (\ref{eq:rss2}) is undefined otherwise at $x_i = y^*$.

We now present the two proposed methods, one by one.

\subsection{Triangulation-based Localization}
\begin{definition}
Let $\alpha_{(Q,D)} = \left(\frac{3}{2\pi D e^{\frac{1}{D}}}\right)^{\frac{3}{2}} Q$. Then (\ref{eq:rss2}) at sensor $i$ becomes:
\begin{equation} \label{e2}
z_i = \frac{\alpha_{(Q,D)}}{d_i^3} + \omega_i.
\end{equation}
\end{definition}
Then, the measurements $z_i, i \in \{1, \cdots,N+1\}$ by the sensors give rise to the following equations:
\begin{equation} \label{e9}
\begin{split}
z_i & = \frac{\alpha_{(Q,D)}}{d_i^3} \Rightarrow\\
 \|x_i - y^*\|^3 &= \frac{\alpha_{(Q,D)}}{z_i}\\
(x_i -y^*)^T(x_i - y^*) & = \left(\frac{\alpha_{(Q,D)}}{z_i}\right)^{\frac{2}{3}}\\
\end{split}
\end{equation}

Taking any two equations from (\ref{e9}), we realise (\ref{e10}).
\begin{equation} \label{e10}
\begin{split}
x_i^T x_i -x_j^T x_j -2(x_i^T - x_j^T) y^* = \left(\frac{\alpha_{(Q,D)}}{z_i}\right)^{\frac{2}{3}}-\left(\frac{\alpha_{(Q,D)}}{z_j}\right)^{\frac{2}{3}}\\
-2(x_i^T - x_j^T) y^* = \left(\frac{\alpha_{(Q,D)}}{z_i}\right)^{\frac{2}{3}}-\left(\frac{\alpha_{(Q,D)}}{z_j}\right)^{\frac{2}{3}}-x_i^T x_i +x_j^T x_j \\
\end{split}
\end{equation}

(\ref{e10}) leads to three equations which are used to form the least-squares solution.
Define: 
\begin{equation}
\hat{\bf A} = 
  \begin{bmatrix} x_1^T - x_2^T \\ \vdots \\ x_1^T - x_{N+1}^T \\ x_2^T - x_3^T\\ \vdots \\x_{N}^T - x_{N+1}^T\end{bmatrix} 
\end{equation}
and 
\begin{equation}
\hat{\bf B} = 
  \begin{bmatrix} \left(\frac{\alpha_{(Q,D)}}{z_1}\right)^{\frac{2}{3}}-\left(\frac{\alpha_{(Q,D)}}{z_2}\right)^{\frac{2}{3}}-x_1^T x_1 +x_2^T x_2 \\ \vdots \\ 
\left(\frac{\alpha_{(Q,D)}}{z_1}\right)^{\frac{2}{3}}-\left(\frac{\alpha_{(Q,D)}}{z_{N+1}}\right)^{\frac{2}{3}}-x_1^T x_1 +x_{N+1}^T x_{N+1} \\
 \left(\frac{\alpha_{(Q,D)}}{z_2}\right)^{\frac{2}{3}}-\left(\frac{\alpha_{(Q,D)}}{z_3}\right)^{\frac{2}{3}}-x_2^T x_2 +x_3^T x_3\\ 
\vdots \\
\left(\frac{\alpha_{(Q,D)}}{z_N}\right)^{\frac{2}{3}}-\left(\frac{\alpha_{(Q,D)}}{z_{N+1}}\right)^{\frac{2}{3}}-x_N^T x_N +x_{N+1}^T x_{N+1}\end{bmatrix} 
\end{equation}

Then, the triangulation based estimate $\hat{y^*}$ of location of the molecular source is given by:

\begin{equation} \label{e11}
\hat{y^*} = \left( \hat{\bf A}^T \hat{\bf A}\right)^{-1} \hat{\bf A}^T \hat{\bf B}
\end{equation}

\subsection{CRB for Triangulation-based Localization}
\label{subsec:crlb}

We now derive the generalized CRB for Triangulation-based localization of the molecular source presented above. Assuming i.i.d. measurements by the sensor nodes, the fusion center constructs the joint probability density function as follows:

\begin{equation} \label{e12}
\begin{split}
f_{(y^*|z_1,\cdots,z_n)} &=\prod_{i=1}^{n} \frac{\left(\frac{\alpha_{(Q,D)}}{d_i^3}\right)^{z_i} e^{-\frac{\alpha_{(Q,D)}}{d_i^3}}}{z_i!}
\end{split}
\end{equation}

The log-likelihood function, $\log{f_{(y^*|z_1,\cdots,z_n)}} =L_{(y^*|z_1,\cdots,z_n)}$ is:

\begin{equation} \label{e13}
\begin{split}
 L_{(y^*|z_1, \cdots, z_n)} &=\sum_{i=1}^{n} z_i \log{\frac{\alpha_{(Q,D)}}{d_i^3}} -\frac{\alpha_{(Q,D)}}{d_i^3} -\log{z_i !}\\
&=-\left(\sum_{i=1}^{n} 3z_i \log{d_i} +\frac{\alpha_{(Q,D)}}{d_i^3}\right) +\kappa_{(z_1,\cdots,z_n)}
\end{split}
\end{equation}

where $\kappa_{(z_1,\cdots,z_n)}$ is a constant. The first and second derivatives with respect to $y^*$ will yield:
 \begin{equation} \label{e14}
\begin{split}
 \dot{L}_{(y^*|z_1, \cdots, z_n)} &=3 \sum_{i=1}^{n} \left(\frac{\alpha_{(Q,D)}}{d_i^4} -\frac{z_i}{d_i} \right) \frac{1}{d_i} \left(x_i -y^* \right)
\end{split}
\end{equation}

 \begin{equation} \label{e15}
\begin{split}
 \ddot{L}_{(y^*|z_1, \cdots, z_n)} &=-3 \sum_{i=1}^{n} \left(\frac{5\alpha_{(Q,D)}}{d_i^7} -2\frac{z_i}{d_i^4} \right)  \left(x_i -y^* \right) \left(x_i -y^* \right)^T \\
&+\frac{1}{d_i^2}\left(\frac{\alpha_{(Q,D)}}{d_i^3} -z_i\right){\bf I}_{n\times n}
\end{split}
\end{equation}

where ${\bf I}_{n\times n}$ is the identity matrix of size $n\times n$. From (\ref{e15}), we get the Fisher information matrix as:

 \begin{equation} \label{e16}
\begin{split}
 -\mathrm{E}\left[\ddot{L}_{(y^*|z_1, \cdots, z_n)}\right] &=9\alpha_{(Q,D)} \sum_{i=1}^{n} \frac{1}{d_i^7}  \left(x_i -y^* \right) \left(x_i -y^* \right)^T 
\end{split}
\end{equation}
where $\mathrm{E}(.)$ is the expectation operator.
The CRB is thus the trace of the inverse of (\ref{e16}).
 \begin{equation} \label{e17}
\begin{split}
 \text{CRB} &=\frac{1}{9\alpha_{(Q,D)} }\text{Tr} \left( \sum_{i=1}^{n} \frac{1}{d_i^7}  \left(x_i -y^* \right) \left(x_i -y^* \right)^T \right)^{-1}.
\end{split}
\end{equation}
where $\text{Tr}(.)$ is the trace of a matrix.

\subsection{Gradient Descent-based Localization}

\begin{lemma} \label{lm:le1}
Under (\ref{eq:rss2}) and assumption \ref{assumption:ass2}, $z_i$ is analytic and a strictly decreasing function of $d_i$ in the noise-free case.
\end{lemma}

{\it Proof.}
We notice that
\begin{equation}
\dot{z_i} = -3\left( \frac{3}{2 \pi D e^{\frac{1}{D}}}\right)^{\frac{3}{2}} \frac{Q}{d_i^4}
\le 0 \; \forall i
\end{equation}
This concludes the proof.

Consequent to lemma \ref{lm:le1}, we can apply the gradient descent minimization procedure to the following non-convex cost function \cite{HBW:SPL:2013}:
\begin{equation}
	\label{eq:J}
	J({y}) = \sum_{i=1}^n (z_i - g(|| d_i ||))^2
\end{equation}
where $g(d_i) = c_{max,i}(d_i) = \frac{\alpha_{(Q,D)}}{d_i^3}$.
Then, the gradient descent control law at fusion center is the following:
\begin{equation}
	\label{eq:gdcl}
	{y}[k+1] = {y}[k] - \mu \frac{\partial J({y})}{\partial {y}} \bigg|_{{y}={y}[k]}
\end{equation}
where $\mu>0$ is the step size of the algorithm. With the knowledge of $Q$, $D$, $z_i$ and ${y}[k]$, (\ref{eq:gdcl}) is implementable at the fusion center. Specifically, the gradient of ${y}$ is given as:
\begin{equation}
	\label{eq:grd_of_y}
	\frac{\partial J({y})}{\partial {y}} = 2 \sum_{i=1}^n \frac{(z_i - g_i(d_i) )\dot{g_i}(d_i) ({x_i} - {y}) } {|| d_i ||}
\end{equation}

It has been shown in \cite{HBW:SPL:2013} that given (\ref{eq:grd_of_y}), suppose there are precisely $n = N+1$ measurement sensors in $\mathbb{R}^N$, and the source $y^*$ is in the open convex hull of the sensor locations $x_i, i \in \{ 1, \cdots, N+1\}$, then: (i) there is a unique point within the open convex hull of the sensor locations where (\ref{eq:grd_of_y}) and (\ref{eq:J}) are identically zero, (ii) the gradient descent law converges uniformly to the true optima, i.e., ${y*}$  in the absence of noise. Notice that this is the same minimum number of sensors required for triangulation-based methods.

\section{Numerical Results}
\label{sec:results}

\begin{figure}[ht]
\begin{center}
	\includegraphics[width=\linewidth]{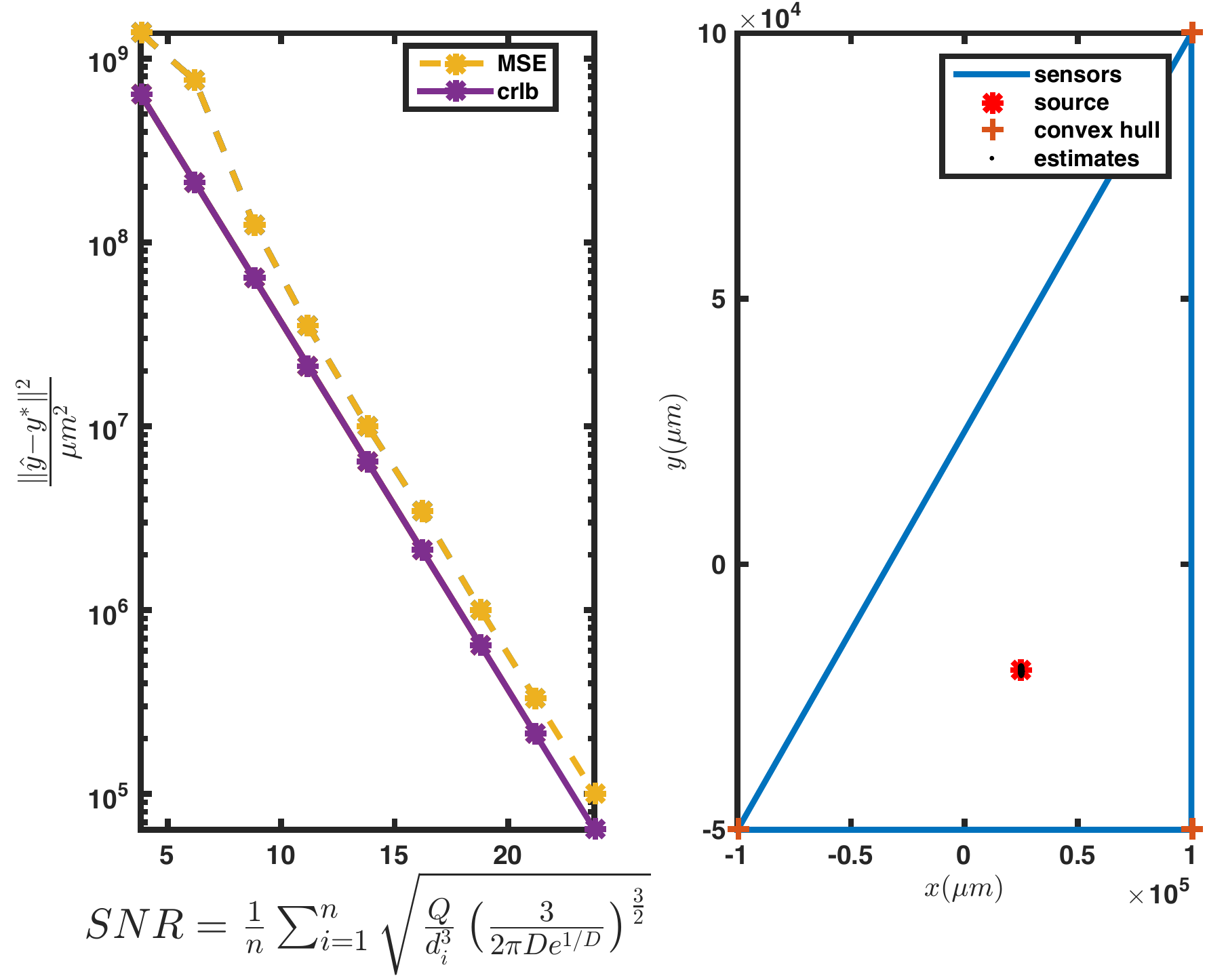} 
\caption{Triangulation-based approach performs very close to the CRB over the whole range of SNR values.}
\label{fig:ls}
\end{center}
\end{figure}

Fig. \ref{fig:ls} investigates the performance of the triangulation-based localization approach. For Fig. \ref{fig:ls} (a), signal-to-noise ratio (SNR) was defined as: $\text{SNR}=\frac{1}{n} \sum_{i=1}^{n} \sqrt{\frac{\alpha_{(Q,D)}}{d_i^3}}$. Fig. \ref{fig:ls} (a) shows that both CRB and MSE decrease with the increase in the SNR, as expected. Additionally, we learn that the mean squared error (MSE) of the triangulation-based approach stays very close to the CRB for all the SNR values. Fig. \ref{fig:ls} (b) plots a 2D layout whereby the molecular source lies within the convex hull of the three sensor nodes. We see that the triangulation based location estimate given by (\ref{e11}) is nearly superimposed on the true location of the source. 

\begin{figure}[ht]
\begin{center}
	\includegraphics[width=3.9in]{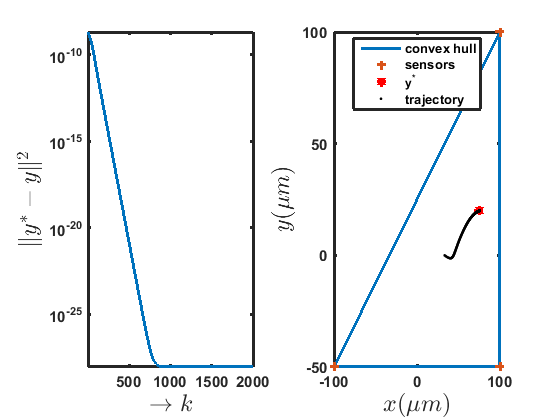} 
\caption{Gradient descent-based approach converges to the true source location uniformly. }
\label{fig:gd}
\end{center}
\end{figure}

Fig. \ref{fig:gd} investigates the performance of the gradient-descent based localization approach. Specifically, Fig. \ref{fig:gd} (a) plots the error $||y^*-y[k]||^2$ against the number of iterations $k$. We learn that the error vanishes, and thus, the gradient descent-based method converges to the true optima/source location uniformly (in less than hundred iterations). Fig. \ref{fig:gd} (b) is again a 2D layout whereby the molecular source lies within the convex hull of the three sensor nodes. Again, from the trajectory of the iterated estimate $y[k]$, we see that the gradient descent method converges to the true location of the source in less than hundred iterations.

\section{Conclusion}
\label{sec:conclusion}

In this paper, we proposed two methods which both utilize (the peak of) the channel impulse response measurements to uniquely localize the molecular source of interest. The first method, the triangulation-based approach, estimates the unknown location of the molecular source using least-squares method. The corresponding CRB was also derived. The second method, basically an iterative approach, utilizes gradient descent law to minimize a non-convex cost function. Simulation results showed that the triangulation-based method performs very close to the CRB, for any given signal-to-noise ratio. Additionally, the gradient descent-based method converges to the true optima/source location uniformly (in less than hundred iterations). 

One potential follow-up work could be to do source localization using time-of-arrival measurements and compare its performance against the two CIR-based localization methods proposed in this work. Another promising direction will be to consider the effect of interference caused by previously sent molecules on the performance of the proposed source localization methods.

Some futuristic applications in the healthcare domain that could potentially benefit from the proposed method include early disease detection, targeted drug delivery, and quick toxicity detection. 


\section*{References}

\bibliography{mybibfile}

\begin{thebibliography}{10}
\expandafter\ifx\csname url\endcsname\relax
  \def\url#1{\texttt{#1}}\fi
\expandafter\ifx\csname urlprefix\endcsname\relax\def\urlprefix{URL }\fi
\expandafter\ifx\csname href\endcsname\relax
  \def\href#1#2{#2} \def\path#1{#1}\fi

\bibitem{Ian:TIT:2013}
M.~Pierobon, I.~F. Akyildiz, Capacity of a diffusion-based molecular
  communication system with channel memory and molecular noise, IEEE
  Transactions on Information Theory 59~(2) (2013) 942--954.
\newblock \href {http://dx.doi.org/10.1109/TIT.2012.2219496}
  {\path{doi:10.1109/TIT.2012.2219496}}.

\bibitem{Nakano:TMBMC:2017}
T.~{Nakano}, Molecular communication: A 10 year retrospective, IEEE
  Transactions on Molecular, Biological and Multi-Scale Communications 3~(2)
  (2017) 71--78.
\newblock \href {http://dx.doi.org/10.1109/TMBMC.2017.2750148}
  {\path{doi:10.1109/TMBMC.2017.2750148}}.

\bibitem{Wang:Chinacomm:2017}
J.~Wang, B.~Yin, M.~Peng, Diffusion based molecular communication: principle,
  key technologies, and challenges, China Communications 14~(2) (2017) 1--18.
\newblock \href {http://dx.doi.org/10.1109/CC.2017.7868158}
  {\path{doi:10.1109/CC.2017.7868158}}.

\bibitem{Ian:CN:2008}
I.~F. Akyildiz, F.~Brunetti, C.~Blázquez,
  \href{http://www.sciencedirect.com/science/article/pii/S1389128608001151}{Nanonetworks:
  A new communication paradigm}, Computer Networks 52~(12) (2008) 2260 -- 2279.
\newblock \href
  {http://dx.doi.org/https://doi.org/10.1016/j.comnet.2008.04.001}
  {\path{doi:https://doi.org/10.1016/j.comnet.2008.04.001}}.
\newline\urlprefix\url{http://www.sciencedirect.com/science/article/pii/S1389128608001151}

\bibitem{Llatser:JSAC:2013}
I.~Llatser, A.~Cabellos-Aparicio, M.~Pierobon, E.~Alarcon, Detection techniques
  for diffusion-based molecular communication, IEEE Journal on Selected Areas
  in Communications 31~(12) (2013) 726--734.
\newblock \href {http://dx.doi.org/10.1109/JSAC.2013.SUP2.1213005}
  {\path{doi:10.1109/JSAC.2013.SUP2.1213005}}.

\bibitem{Kuran:ICC:2011}
M.~S. Kuran, H.~B. Yilmaz, T.~Tugcu, I.~F. Akyildiz, Modulation techniques for
  communication via diffusion in nanonetworks, in: Communications (ICC), 2011
  IEEE International Conference on, IEEE, 2011, pp. 1--5.

\bibitem{Sayed:SPMag:2005}
A.~Sayed, A.~Tarighat, N.~Khajehnouri, Network-based wireless location:
  challenges faced in developing techniques for accurate wireless location
  information, Signal Processing Magazine, IEEE 22~(4) (2005) 24--40.
\newblock \href {http://dx.doi.org/10.1109/MSP.2005.1458275}
  {\path{doi:10.1109/MSP.2005.1458275}}.

\bibitem{Armstrong:CommMag:2013}
J.~Armstrong, Y.~A. Sekercioglu, A.~Neild, Visible light positioning: a roadmap
  for international standardization, IEEE Communications Magazine 51~(12)
  (2013) 68--73.
\newblock \href {http://dx.doi.org/10.1109/MCOM.2013.6685759}
  {\path{doi:10.1109/MCOM.2013.6685759}}.

\bibitem{KUNDU:Ultrasonics4:2014}
T.~Kundu,
  \href{http://www.sciencedirect.com/science/article/pii/S0041624X13001819}{Acoustic
  source localization}, Ultrasonics 54~(1) (2014) 25 -- 38.
\newblock \href
  {http://dx.doi.org/https://doi.org/10.1016/j.ultras.2013.06.009}
  {\path{doi:https://doi.org/10.1016/j.ultras.2013.06.009}}.
\newline\urlprefix\url{http://www.sciencedirect.com/science/article/pii/S0041624X13001819}

\bibitem{HBW:SPL:2013}
H.~Baidoo-Williams, S.~Dasgupta, R.~Mudumbai, E.~Bai, On the gradient descent
  localization of radioactive sources, Signal Processing Letters, IEEE 20~(11)
  (2013) 1046--1049.
\newblock \href {http://dx.doi.org/10.1109/LSP.2013.2279499}
  {\path{doi:10.1109/LSP.2013.2279499}}.

\bibitem{Dasgupta:SPL:2008}
C.~Meng, Z.~Ding, S.~Dasgupta, A semidefinite programming approach to source
  localization in wireless sensor networks, Signal Processing Letters, IEEE 15
  (2008) 253--256.
\newblock \href {http://dx.doi.org/10.1109/LSP.2008.916731}
  {\path{doi:10.1109/LSP.2008.916731}}.

\bibitem{Dasgupta:TSP:2011}
E.~Xu, Z.~Ding, S.~Dasgupta, Source localization in wireless sensor networks
  from signal time-of-arrival measurements, Signal Processing, IEEE
  Transactions on 59~(6) (2011) 2887--2897.
\newblock \href {http://dx.doi.org/10.1109/TSP.2011.2116012}
  {\path{doi:10.1109/TSP.2011.2116012}}.

\bibitem{Moore:TSP:2012}
M.~Moore, T.~Nakano, A.~Enomoto, T.~Suda, Measuring distance from single spike
  feedback signals in molecular communication, Signal Processing, IEEE
  Transactions on 60~(7) (2012) 3576--3587.
\newblock \href {http://dx.doi.org/10.1109/TSP.2012.2193571}
  {\path{doi:10.1109/TSP.2012.2193571}}.

\bibitem{Wang:CommL:2015}
X.~Wang, M.~D. Higgins, M.~S. Leeson, Distance estimation schemes for diffusion
  based molecular communication systems, IEEE Communications Letters 19~(3)
  (2015) 399--402.
\newblock \href {http://dx.doi.org/10.1109/LCOMM.2014.2387826}
  {\path{doi:10.1109/LCOMM.2014.2387826}}.

\bibitem{Huang:GC:2013}
J.-T. Huang, H.-Y. Lai, Y.~Lee, C.~Lee, P.~Yeh, Distance estimation in
  concentration-based molecular communications, in: 2013 IEEE Global
  Communications Conference (GLOBECOM), 2013, pp. 2587--2591.
\newblock \href {http://dx.doi.org/10.1109/GLOCOM.2013.6831464}
  {\path{doi:10.1109/GLOCOM.2013.6831464}}.

\bibitem{Noel:GC:2014}
A.~Noel, K.~C. Cheung, R.~Schober, Bounds on distance estimation via diffusive
  molecular communication, in: 2014 IEEE Global Communications Conference,
  2014, pp. 2813--2819.
\newblock \href {http://dx.doi.org/10.1109/GLOCOM.2014.7037234}
  {\path{doi:10.1109/GLOCOM.2014.7037234}}.

\bibitem{Nakano:ICC:2014}
Y.~{Okaie}, T.~{Nakano}, T.~{Hara}, K.~{Hosoda}, Y.~{Hiraoka}, S.~{Nishio},
  Modeling and performance evaluation of mobile bionanosensor networks for
  target tracking, in: 2014 IEEE International Conference on Communications
  (ICC), 2014, pp. 3969--3974.
\newblock \href {http://dx.doi.org/10.1109/ICC.2014.6883941}
  {\path{doi:10.1109/ICC.2014.6883941}}.

\bibitem{Okaie:TNB:2014}
Y.~Okaie, T.~Nakano, T.~Hara, T.~Obuchi, K.~Hosoda, Y.~Hiraoka, S.~Nishio,
  Cooperative target tracking by a mobile bionanosensor network, IEEE
  Transactions on NanoBioscience 13~(3) (2014) 267--277.
\newblock \href {http://dx.doi.org/10.1109/TNB.2014.2343237}
  {\path{doi:10.1109/TNB.2014.2343237}}.

\bibitem{Okaie:ICNCC:2015}
Y.~Okaie, T.~Obuchi, T.~Hara, S.~Nishio, In silico experiments of mobile
  bionanosensor networks for target tracking, in: Proceedings of the Second
  Annual International Conference on Nanoscale Computing and Communication,
  ACM, 2015, p.~14.

\bibitem{Iwasaki:Globecom:2016}
S.~{Iwasaki}, J.~{Yang}, A.~O. {Abraham}, J.~L. {Hagad}, T.~{Obuchi},
  T.~{Nakano}, Modeling multi-target detection and gravitation by intelligent
  self-organizing bioparticles, in: 2016 IEEE Global Communications Conference
  (GLOBECOM), 2016, pp. 1--6.
\newblock \href {http://dx.doi.org/10.1109/GLOCOM.2016.7842000}
  {\path{doi:10.1109/GLOCOM.2016.7842000}}.

\bibitem{Nakano:SPAWC:2016}
a.~{Nakano}, a.~{Kobayashi}, a.~{Koujin}, a.~{Chen-Hao Chan}, a.~{Yu-Hsiang
  Hsu}, a.~{Okaie}, a.~{Obuchi}, a.~{Hara}, a.~{Hiraoka}, a.~{Haraguchi},
  Leader-follower based target detection model for mobile molecular
  communication networks, in: 2016 IEEE 17th International Workshop on Signal
  Processing Advances in Wireless Communications (SPAWC), 2016, pp. 1--5.
\newblock \href {http://dx.doi.org/10.1109/SPAWC.2016.7536831}
  {\path{doi:10.1109/SPAWC.2016.7536831}}.

\bibitem{Nakano:TC:2017}
T.~{Nakano}, Y.~{Okaie}, S.~{Kobayashi}, T.~{Koujin}, C.~{Chan}, Y.~{Hsu},
  T.~{Obuchi}, T.~{Hara}, Y.~{Hiraoka}, T.~{Haraguchi}, Performance evaluation
  of leader–follower-based mobile molecular communication networks for target
  detection applications, IEEE Transactions on Communications 65~(2) (2017)
  663--676.
\newblock \href {http://dx.doi.org/10.1109/TCOMM.2016.2628037}
  {\path{doi:10.1109/TCOMM.2016.2628037}}.

\bibitem{Yang:ICSCN:2018}
L.~{Yang}, Y.~{Mao}, Q.~{Liu}, H.~{Zhai}, K.~{Yang}, High-efficiency target
  detection scheme through relay nodes in chemotactic-based molecular
  communication, in: 2018 IEEE International Conference on Sensing,
  Communication and Networking (SECON Workshops), 2018, pp. 1--4.
\newblock \href {http://dx.doi.org/10.1109/SECONW.2018.8396349}
  {\path{doi:10.1109/SECONW.2018.8396349}}.

\bibitem{Giaretta:TIFS:2016}
A.~{Giaretta}, S.~{Balasubramaniam}, M.~{Conti}, Security vulnerabilities and
  countermeasures for target localization in bio-nanothings communication
  networks, IEEE Transactions on Information Forensics and Security 11~(4)
  (2016) 665--676.
\newblock \href {http://dx.doi.org/10.1109/TIFS.2015.2505632}
  {\path{doi:10.1109/TIFS.2015.2505632}}.

\bibitem{Raz:TNB:2015}
N.~R. Raz, M.~Akbarzadeh-T.*, M.~Tafaghodi, Bioinspired nanonetworks for
  targeted cancer drug delivery, IEEE Transactions on NanoBioscience 14~(8)
  (2015) 894--906.
\newblock \href {http://dx.doi.org/10.1109/TNB.2015.2489761}
  {\path{doi:10.1109/TNB.2015.2489761}}.

\bibitem{Noel:Globecom:2016}
A.~Noel, Y.~Deng, D.~Makrakis, A.~Hafid, Active versus passive: Receiver model
  transforms for diffusive molecular communication, in: 2016 IEEE Global
  Communications Conference (GLOBECOM), IEEE, 2016, pp. 1--6.

\end{thebibliography}

\end{document}